\documentclass[nofootinbib,aps]{revtex4}

\usepackage{amssymb,amstext,amsmath}
\usepackage[dvips]{graphicx}
\usepackage{latexsym}
\usepackage{psfrag}
\usepackage{amsfonts}
\usepackage{bbm}
\usepackage{axodraw}
\usepackage{color}

\newcommand{\Ref}[1]{(\ref{#1})}

\newcommand{\eqa}{\begin{eqnarray}}
\newcommand{\neqa}{\end{eqnarray}}
\newcommand{\equ}{\begin{equation}}
\newcommand{\nequ}{\end{equation}}

\def\om{\omega}

\def\la{\langle}
\def\ra{\rangle}
\newcommand{\bra}[1]{\la {#1}|}
\newcommand{\ket}[1]{|{#1}\ra}
\newcommand{\mean}[1]{\la{#1}\ra}

\def\d{\delta}
\def\f{\frac}

\newcommand{\p}{\partial}

\usepackage{bbm}

\newcommand{\scr}{\rm\scriptscriptstyle}

\newcommand{\SU}{\mathrm{SU}}

\newcommand{\lp}{\ell_{\rm P}}

\begin{document}

\title{\Large\bf Towards the graviton from spinfoams: \\ the 3d toy model}
\author{{\bf Simone Speziale}\footnote{sspeziale@perimeterinstitute.ca}}
\affiliation{Perimeter Institute, 31 Caroline St. N, Waterloo, N2L 2Y5,
Ontario, Canada.}

\begin{abstract}{
\noindent {Recently, a proposal has appeared for the extraction of the 2-point function
of linearised quantum gravity, 
within the spinfoam formalism.
This relies on the use of a boundary state, which introduces a semi--classical flat geometry on the
boundary.
In this paper, we investigate this proposal considering a toy
model in the (Riemannian) 3d case, where the
semi--classical limit is better understood. We show that in this limit 
the propagation kernel of the model is the one for the harmonic oscillator. This is at the origin
of the expected $1/\ell$ behaviour of the 2-point function. Furthermore, we numerically study
the short scales regime, where deviations from this behaviour occur.}}
\end{abstract}

\maketitle

\section{Introduction}
The spinfoam formalism \cite{carlo} is a candidate covariant approach to a non--perturbative
quantisation of General Relativity (GR).  At present, it seems to provide a consistent
background independent theory at the Planck scale, where it describes a (discrete)
quantum geometry. However, the large scale behaviour is less understood.
In particular, the formalism lacks a well--defined procedure to study the semi--classical limit, 
define particle scattering amplitudes and reproduce low--energy physics.
Consider the pure gravity case: in the low--energy limit, we would expect to recover the conventional perturbative expansion described
in terms of gravitons. The importance of obtaining this would be two--fold:
on the one hand, it would provide fundamental ground for
the emergence of General Relativity in the large scale limit of spinfoams,
thus proving it to be a sensible theory of quantum gravity.
On the other hand, it would open the way to the exploration of the
corrections to the propagation of gravitons due to the microscopical
quantum geometry, where new physics is potentially expected.

There is a simple heuristic reason that shows the difficulty to recover 
conventional quantum field theory (QFT) from background independent approaches:
the arguments of the $n$-point functions used in QFT are 
spacetime point coordinates, which are not defined in the absence of a background.
Recently, a possible way out of this problem has been suggested in \cite{ModestoProp}.
The key idea is to construct $n$-point functions
by means of the propagation kernel, which provides an amplitude
for the fields assigned on a boundary of spacetime. Since the
value of the gravitational field provides the boundary with a metric,
$n$-point functions can be defined with respect to this metric.
In the deep quantum gravity regime, one is willing to consider
arbitrary boundaries \cite{Conrady, Oeckl}. However, to recover the behaviour
of linearised quantum gravity, it seems as a first step sufficient to simply consider the
usual setting of QFT, when the boundary is given by two spacelike hyperplanes,
and the boundary values of the gravitational field are small perturbations over
flat space, $h_{ab}'$ and $h_{ab}''$, which are asymptotically vanishing. 
In this context, the propagation kernel $W[h', h'', T]$ for
linearised GR was considered in \cite{Kuchar, teit2}, and explicitly derived in 
the temporal gauge in \cite{Mattei}.
The time $T$ entering the kernel is the asymptotic proper time between the two
hyperplanes. Indeed, it is a simple exercise to recover 
for instance the $2$-point function $W_{abcd}(x, y)$
by sandwiching $W[h', h'', T]$ with two one--particle states $\Psi_1[h']_{ab}(x)$,
$\Psi_1[h'']_{cd}(y)$, and integrating over the boundary values $h_{ab}'$ and $h_{ab}''$:
\eqa\label{WGR}
W_{abcd}(\vec x, \vec y; T) &=& \f1{\cal N} \ \int {\cal D}h' \ {\cal D}h'' \ W[h', h'', T]\ \Psi_0[h'] 
\ h'_{ab}(0, \vec x) \ \Psi_0[h''] \ h''_{cd}(T, \vec y),
\neqa
where the normalisation constant $\cal N$ is given by the functional integral without
the field insertions.\footnote{Strictly speaking, this is only the advanced Green function $\Delta_+(x,y)$.
The Feynman propagator is obtained by taking 
$\theta(T)\Delta_+(x,y)+\theta(-T)\Delta_-(x,y)$.}
Notice that both the kernel $W$ and the vacuum state $\Psi_0$ (encoded in the kernel, see \cite{Mattei})
are gauge invariant expressions; however, the field insertions
$h'_{ab}$ and $h''_{cd}$ are not gauge invariant, 
thus the evaluation of the expression above requires an additional spatial gauge--fixing.
Choosing for instance the Coulomb--like gauge $\p_a h_{ab}=0$, we have
\equ\label{WGR1}
W_{abcd}(\vec x, \vec y; T) =
- \ell_{\rm P} \int \f{d^{n-1}p}{(2\pi)^{n-1}} \ \f{e^{-i p (x-y)}}{2\om_p} \ 
\Big(D_{ac}D_{bd} + D_{ad}D_{bc} - \f2{n-2} D_{ab}D_{cd}\Big),
\nequ
where $n$ is the  dimension of spacetime, and $D_{ab}=\d_{ab}-\f{p_a p_b}{p^2}$ is the
transverse projector. Different gauge choices result in different numerical factors in the tensorial
structure.

Following \cite{ModestoProp}, the 2-point function of linearised quantum gravity
\Ref{WGR} can be translated into the spinfoam formalism using the expression
\equ\label{W1}
W_{\mu\nu\rho\sigma}(x, y) = \f1{\cal N} \ \sum_{s} \ W[s]\ \Psi_0[s] 
\ \bra{s}h_{\mu\nu}(x)\ket{s} \ \bra{s}h_{\rho\sigma}(y)\ket{s}.
\nequ
Here the sum is over spin networks $s$; the propagation
kernel $W[s]$ is provided by the explicit spinfoam model chosen, 
and it includes a sum over all spinfoams $\sigma$ compatible with $s$.
The boundary state $\Psi_0[s]$ is a weighted functional of spin networks
peaked around those reproducing flat space in the semi--classical limit.

This approach has been considered in \cite{RovelliProp} for a proposal to extract
the graviton from the (Riemannian) 4d Barrett--Crane model. The results there described
are promising; indeed, considering a single component of the 2-point
function, one does recover the correct $1/\ell^{2}$ dependence on the distance. 
However, there are many issues that still needs to be clarified. In particular,
the role played by the Gaussian boundary state used to introduce the flat
geometry on the boundary, and the way this geometry determines the distance dependence.

Remarkably, these issues are present in a  very similar way also in the 3d case.
The 3d case has the advantage that the semi--classical limit is better understood,
and there are no degenerate configurations plaguing the limit.
In this paper, we propose to apply the strategy of \cite{RovelliProp} in 3d, in the hope to 
clarify some of the issues.
As it is well known, 3d GR has no local degrees of freedom. Nonetheless, the
2-point function of the linearised quantum theory is a well defined quantity
that can be evaluated once a gauge--fixing is chosen. For instance, 
it is given in the Coulomb gauge by \Ref{WGR1} with $n=3$; it shows the usual
$1/p^2$ behaviour, which in 3d means a dependence $1/\ell$
over the spacetime distance. However, this quantity is a pure gauge, thus the
quantum theory does not have a propagating graviton \cite{Deser}.

Notice also that in 3d the Newton constant $G$ has inverse mass dimensions (in units $c=1$);
we define the 3d Planck length as $\ell_{\rm P}=16\pi \hbar G$.

\section{The toy model}
The toy model we consider here is a tetrahedron with dynamics described by the
Regge action, whose fundamental variables are the edge lengths $\ell_e$. 
Since we are dealing with a single tetrahedron, all edges are boundary edges,
and the action consists only of the boundary term, namely it coincides with the Hamilton function
of the system:
\equ\label{regge}
S_{\rm R}[\ell_e] = \f1{16\pi G} \sum_{e}\, \ell_e \, \theta_e(\ell_e).
\nequ
Here the $\theta_e$ are the dihedral angles  of the tetrahedron, namely the angles between
the outward normals to the triangles. They represent a discrete version of the extrinsic curvature \cite{Sorkin}.
Furthermore, they satisfy the non--trivial relation
\equ\label{deficit}
{16\pi G} \, \f{\p S_{\rm R}}{\p\ell_e} =  \theta_e.
\nequ

In this discrete setting, assigning the six edge lengths is equivalent
to the assignment of the boundary gravitational field.
In particular, let us consider the case when two opposite edges of the 
tetrahedron have lengths $a$ and $b$, and the remaining four edges have all
the same length, say $c$. Following \cite{nutshell}, we know that the Hamilton
function $S[a, b, c]$ given by \Ref{regge} defines a simple relativistic system,
of which $a$, $b$ and $c$ are partial observables \cite{RovelliPartial}. That is,
they include both the independent (time) variable, and the dependent (dynamical)
variables, and the dynamics provides a relation amongst them. 
In the following, we choose $c$ as the time variable. 
The classical dynamics of the system is described using the Hamilton
function to read the evolution of $b$ in a time $c$, given the initial value $a$.

The quantum dynamics, on the other hand, is
described by the Ponzano--Regge (PR) model \cite{Ponzano}. In the model, the lengths
are promoted to operators $\lp^2 X_e^IX_e^I$, whose spectrum is given by $\ell_e^2 = \ell_{\rm P}^2 C^2(j_e)$,
where the half--integer $j$ labels $\SU(2)$ irreducible representations
(irreps), and $C^2(j)=j(j+1)+\f14$ is the Casimir operator.\footnote{Conventionally,
the $\SU(2)$ Casimir is taken to be simply $j(j+1)$. However, it is defined up to 
an additive constant, and the value $1/4$ chosen here allows a matching with the
original PR ansatz, $\ell_e=\lp (j_e+\f12)$.}
In the model, each tetrahedron has an amplitude given by
Wigner's $\{6j\}$ symbol for the recoupling theory of $\SU(2)$.
Under the identifications $a=\ell_{\rm P} C(j_1)$, $b=\ell_{\rm P} C(j_2)$,
$c=\ell_{\rm P} C(j_0)$, the propagation kernel for our toy model is
\equ\label{W}
W[j_1, j_2, j_0] = \prod_e \, (2 j_e +1) \; \left\{ 
\begin{array}{ccc} j_1 & j_0 & j_0 \\ j_2 & j_0 & j_0 \end{array} \right\}.
\nequ
It represents the amplitude for the (eigen)values of the boundary gravitational field.
Notice that we can assume to have measured the quantity $j_0$, and then interpret
$W[j_1, j_2, j_0]$ as the transition amplitude between $j_1$ and $j_2$ in a time $j_0$.

The key property of the PR model which we will need in the following
is that in the large $j$ regime we have \cite{Ponzano, asympt, asympt2}
\equ\label{asymp}
\lim_{j\mapsto\infty} W[j_1, j_2, j_0]  \sim
\prod_e \; (2j_e+1) \;
\sqrt{\frac{2}{3\pi V(j_e)}}\;
\cos\left( \sum_e (j_e +\f12)\theta_e(j_e) +\f\pi 4  \right),
\nequ
where the argument of the cosine can be recognised to be
($1/\hbar$ times) the Regge action \Ref{regge}, with
$\ell_e=\lp C(j_e)$ (the constant factor $\frac{\pi}{4}$ appearing in~\Ref{asymp} 
clearly does not affect the classical dynamics).
Usually, we would expect the semi--classical limit to be the exponential of the
classical action. However, the presence of the two exponentials giving rise
to the cosine can be understood as follows.
The sign in front of the action is determined by the orientation
of the manifold; it is reversed if we invert the orientation of $M$.
However, the classical theory does not
distinguish between the two orientation, thus the path integral sums over both.
We will see below that only one contributes to the propagator.
Therefore, the model admits a clear semi--classical limit, obtained by taking
$j\mapsto\infty$, $\ell_{\rm P} \mapsto 0$, such that the product ${\lp}\,j$
is finite.

\section{The boundary state}
\begin{figure}[ht]
\begin{center}\includegraphics[width=5cm]{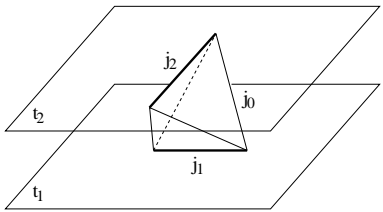}
\end{center}\caption{\small{The dynamical tetrahedron as evolution
between two hyperplanes. The labels give the physical lengths as
$a=\lp C(j_1)$, $b=\lp C(j_2)$, $c=\lp C(j_0)$, and 
$T=t_2-t_1=c/\sqrt 2$.}}\label{planes}
\end{figure}

To make a link between this toy model and the 2-point function of GR, we consider 
two hyperplanes in 3d Euclidean spacetime, and we embed the tetrahedron as in
Fig.\ref{planes}.
Namely, we assume that we have measured the time interval $T:=t_2-t_1$,
and we compute the correlator between perturbations of the length of the bottom edge, say $\ell_a$,
and perturbations of the top edge, $\ell_b$. 
Because the two edges are embedded in flat space, we can assume 
the coordinate vector $\ell^\mu_a$ to be along
the $x^1$ direction and $\ell^\mu_b$ along the $x^2$ direction, as in Fig.\ref{planes}.
With this setting, we are naturally led to consider the projection 
$\ell_a^\mu(x) \ell_a^\nu(x) \ell_b^\rho(y) \ell_b^\sigma(y) W_{\mu\nu\rho\sigma}(x, y)$
of \Ref{W1},
where $x$ and $y$ are the mid--points respectively of $\ell_a$ and $\ell_b$; therefore
we restrict our attention to the following quantity,
\equ\label{W1122}
\ell_a^2 \, \ell_b^2 \, W_{1122}(T):= 
\ell_a^\mu(x) \ell_a^\nu(x) \ell_b^\rho(y) \ell_b^\sigma(y) W_{\mu\nu\rho\sigma}(x, y) = 
\ell_a^2 \, \ell_b^2 \,f_\lambda \lp \int \f{d^2p}{(2\pi)^2} \f{e^{i\om_p T}}{2\om_p},
\nequ
where $f_\lambda$ is a numerical factor depending on the choice of gauge (in \Ref{WGR1},
$f_\lambda=2$).

In this ``minimalistic'' toy model, the boundary spin network $s$ appearing in \Ref{W1}
is the boundary spin network of the tetrahedron. Since this is a tetrahedral graph with
links in one--to--one correspondence with the edges of the tetrahedron, we can simply
use $j_e$ for the spin network labels. However, we stress that we are considering a situation
where we have measured the ``time'' variable $T$, thus the labels for the edges in the bulk
are fixed to the value $j_0$.\footnote{Consequently the conjugate variables, namely
the bulk dihedral angles, are maximally spread. Notice that this is different from the setting of
\cite{RovelliProp}, where all labels are allowed to fluctuate. However, a link between the two settings
can be made by considering in \cite{RovelliProp} a Gaussian with different widths for the bulk and
the top and bottom edges, and then studying the limit case when the former goes to zero.} 
With a little trigonometry, we have 
$T= c/{\sqrt 2}$.
Therefore, the sum over $s$ in \Ref{W1} is truly
simply a sum over $j_1$ and $j_2$, and \Ref{W} is the propagation kernel to be used,
\equ\label{W2}
\ell_a^2 \, \ell_b^2 \, W_{1122}(x,y) = \f1{\cal N} \ \sum_{j_1, j_2} \ W[j_1, j_2, j_0]\ \Psi_0[j_1, j_2] 
\ \bra{j_1}\ell^\mu_a \ell^\nu_a h_{\mu\nu}(x)\ket{j_1} \ 
\bra{j_2} \ell^\rho_b \ell^\sigma_b h_{\rho\sigma}(y)\ket{j_2}.
\nequ

The next ingredient entering \Ref{W1} is the vacuum boundary state, $\Psi_0[j_1, j_2]$.
Its role is to peak the boundary spin networks around those
reproducing flat spacetime.
The natural choice is to take a Gaussian around a background value for the variables $j_1$ and $j_2$. 
With respect to this, notice that any value of $j_1$ and $j_2$ is compatible with a flat tetrahedron. 
For simplicity, let us choose an equilateral tetrahedron:
since the bulk edges are fixed to be $j_0$ by the time measurement, this means
peaking the Gaussian around the value $j_0$ for both $j_1$ and $j_2$.
Furthermore, a key observation  in \cite{RovelliProp} is that the 
Gaussian should be peaked on the intrinsic \emph{as well as} the extrinsic geometry
of the boundary.\footnote{This is analogous to a Gaussian representing a coherent state
for a free particle in QM, which is picked on both the position and momentum.}
While the intrinsic classical geometry is represented by the edge lengths, 
the extrinsic geometry, on the other hand, must 
be fixed giving the boundary dihedral angles $\theta_e$. 
For the equilateral tetrahedron, the value of all dihedral angles is $\vartheta = \arccos(-\f{1}{3})$.
Using $j_0$ and $\vartheta$, we consider the tentative boundary state
\equ\label{3dGaussian}
\Psi_0[j_1, j_2] = \f1{{\cal N}_0}\exp \Big\{-\f{\alpha}{2}\sum_{i=1}^2 (j_i - j_0)^2
+ i \vartheta \sum_{i=1}^2 (j_i+\f12) \Big\}.
\nequ
Here $\alpha$ is the width of the Gaussian, which we leave as a free parameter for the moment.
However, let us notice the following. In order for \Ref{3dGaussian} to 
represent a good vacuum boundary state, we require that in the semi--classical
limit all relative uncertainties vanish, that is
\equ\label{limit}
\f{\mean{\Delta j}}{\mean{j}}=\f{1}{j_0\sqrt\alpha} \mapsto 0, 
\qquad \f{\mean{\Delta \theta}}{\mean\theta}=\f{\sqrt \alpha}{\vartheta} \mapsto 0.
\nequ
The semi--classical limit is obtained by taking $j_0\mapsto\infty$.
Since $\vartheta$ is not affected by this limit, the conditions above
imply $\alpha\propto j_0^{-r}$, with $0<r<2$.
In spite of its simplicity, we show below that this Gaussian is
enough to characterize the usual properties of the propagator over flat spacetime.

\section{The field insertions}\label{field}
 
The last ingredients entering \Ref{W2} are the expectation values of the gravitational operator,
such as $\bra{j_1}\ell^\mu_a \ell^\nu_a h_{\mu\nu}(x)\ket{j_1}$.
A crucial point here concerns gauge invariance. 
Let us go back to \Ref{WGR}. As discussed in the introduction, the field insertions are not
gauge invariant quantities, therefore an additional gauge--fixing is required, in order
for \Ref{WGR} to be well defined. Precisely in the same way, a gauge--fixing has to be chosen
to give a well defined prescription to identify $\bra{j_1}\ell^\mu_a \ell^\nu_a h_{\mu\nu}(x)\ket{j_1}$
in the PR model, and thus to evaluate \Ref{W2}.

In particular, notice that we can choose in the continuum the gauge $\ell^\mu_a \ell^\nu_a h_{\mu\nu}(x)=0$;
to be consistent with this choice, we have to set $\bra{j_1}\ell^\mu_a \ell^\nu_a h_{\mu\nu}(x)\ket{j_1}=0$
in the PR model. Consequently, \Ref{W2} is identically zero. The peculiarity of the 3d case is that this choice
of gauge can be made simultaneously for all independent directions:
the 3d graviton propagator is a pure gauge, and
 accordingly all the components can be set to zero. In the 4d case, on the other hand, 
the gauge choice $\ell^\mu_a \ell^\nu_a h_{\mu\nu}(x)=0$ cannot be made for all independent
directions; in the end two components survive, the two degrees of freedom of a spin-2 
massless particle.
However, in the construction presented here we consider a single component of the graviton,
thus we can not give a full proof of this peculiarity. To pursue this further,
we need to extend the model to more tetrahedra, so that we can study the full tensorial
structure.
In fact, once we have the full tensorial structure, the counting of the number of physical degrees
of freedom can be done using the Ward identities, which
relate different diagrams through the symmetries of the theory.
The relevant symmetry of 3d GR is the invariance under diffeomorphisms, which in the
PR model is implemented by the so--called Pachner moves (see for instance \cite{Freidel1}):
these moves involve changing the number of tetrahedra in the triangulation.
Therefore, in order to apply these moves and study the symmetries of the 2-point function,
we have to extend the model to more tetrahedra, and we leave this issue
open for further work.

To study the large scale behaviour emerging from \Ref{W2}, we now proceed choosing a 
Coulomb--like gauge in which
$h_{\mu\nu}$ has components
along the edge $\ell_a$, so that its projections can be identified with non--trivial
PR operators, such as $X_a^I X_a^I$, the length operator introduced above.
More precisely, since $h_{\mu\nu}$ is a small perturbation around flat spacetime, we can write 
$\ell_a^\mu \ell_a^\nu h_{\mu\nu}(x) =\ell_a^\mu \ell_a^\nu (g_{\mu\nu}(x)-\d_{\mu\nu}):= 
\lp^2 X_a^I X_a^I - \lp^2 C^2(j_0)$,
where $\lp^2 C^2(j_0)$ is the background value of the length. Recall that 
$X_a^I X_a^I$ acts diagonally on an edge labeled by $j$, giving $C^2(j)$;
therefore, in this gauge, the semi--classical limit $j_0\mapsto \infty$ gives
\equ\label{insertions}
\bra{j_1}\ell_a^\mu \ell_a^\nu h_{\mu\nu}(x)\ket{j_1}=
\lp^2 [C^2(j_1) - C^2(j_0)] \sim 2j_0 (j_1-j_0)\lp. 
\nequ

\section{Graviton propagator}
We are now ready to perform the evaluation of \Ref{W2}.
To shorten our notation, let us define $\d j_i:=j_i-j_0$, $i=1, 2$.
Using the explicit expressions \Ref{asymp}, \Ref{3dGaussian} and \Ref{insertions},
we have
\equ\label{Wexact}
\ell_a^2 \, \ell_b^2 \, W_{1122}(T)
= \f{\lp^4}{\cal N}\, \sum_{j_1, j_2} \prod_e \, (2j_e+1) \; \{ 6j \} \;
e^{-\f{\alpha}{2}\sum_{i} \d j_i^2 + i \vartheta \sum_i (j_i+\f12) }
\;[C^2(j_1) - C^2(j_0)] \, [C^2(j_2) - C^2(j_0)].
\nequ
Before proceeding, let us stress the logic: we introduced a flat boundary state,
an equilateral tetrahedron with label $j_0$; this enters
into the choice of the Gaussian as well as into the definition of the field insertions.
Then, to have a semi--classical behaviour of its geometry, we consider the large $j_0$ regime. 
In this regime, \Ref{Wexact} is approximated by
\eqa\label{3dW1}
\ell_a^2 \, \ell_b^2 \, W_{1122}(T)
\simeq \f{\lp^4}{\cal N}\; 
 \sum_{j_1, j_2}  \prod_e \, (2 j_e+1) \; 
\f {4j_0^2 \,\d j_1 \, \d j_2}
{\sqrt{6\pi V(j_e)}}
\Big( e^{i S_{\rm R}[j_e]+ i\f{\pi}{4}} + e^{-i S_{\rm R}[j_e] - i\f{\pi}{4}}\Big) \;
e^{-\f{\alpha}{2}\sum_i \d j_i^2+ i \vartheta \sum_i (j_i+\f12) }.
\neqa
This is the expression we now evaluate analytically to study the asymptotic behaviour 
of the 2-point function, which is to be compared with \Ref{W1122}. As in \cite{RovelliProp}, the Gaussian
implies $\d j_i\ll 1$ in the sum, thus we can expand
the Regge action around $j_0$,
\equ\label{ReggeE}
S_{\rm R}[j_e]=\sum_e (j_e + \f12)\theta_e(j_e) \simeq 
\sum_{e} \,(j_0+\f12) \, \vartheta + \sum_i\f{\p S_{\rm R}}{\p j_i}|_{j_e=j_0} \d j_i +
\f{1}{2} \sum_{i,k} \f{\p^2 S_{\rm R}}{\p j_i \p j_{k}}|_{j_e=j_0} \d j_i \d j_{k}.
\nequ
Using \Ref{deficit}, we have $\f{\p S_{\rm R}}{\p j_i}|_{j_e=j_0}=\vartheta$.
Introducing $G_{ik} = \f{\p \theta_i}{\p j_{k}}|_{j_e=j_0}$, we can write
\equ\label{ReggeExp}
S_{\rm R}[j_e] \simeq 4\,\vartheta\, (j_0+\f12)+
\vartheta \sum_{i=1}^2 \,(j_i+\f12) + \f{1}{2} \sum_{i,k=1}^2 G_{ik} \d j_i \d j_{k}.
\nequ
$G_{ik}$ can be calculated from elementary geometry; for the equilateral tetrahedron at hand,
$G_{11}=G_{22}=-\f{\sqrt 2}{3 j_0}$, $G_{12}=G_{21}=-\f{\sqrt 2}{j_0}$.
If we insert this expansion into \Ref{3dW1}, the two exponentials become
$$
e^{i\f{\pi}{4} + 4i \vartheta(j_0+\f12)+
2i\vartheta \sum_{i} \,(j_i+\f12) + \f{i}{2} \sum_{i,k} G_{ik} \d j_i \d j_{k} 
-\f{\alpha}{2}\sum_{i} \d j_i^2}
+ e^{-i\f{\pi}{4} - 4i \vartheta (j_0+\f12) - \f{i}{2} \sum_{i,k} G_{ik} \d j_i \d j_{k} -\f{\alpha}{2}\sum_{i} \d j_i^2}.
$$
The first one is a rapidly oscillating term in $j_i$, which vanishes when we perform the sum \Ref{3dW1},
so that we can consider only the second term. Furthermore, notice that 
at this order in the expansion in $\d j_i$, we have $V(j_e)\sim V(j_0)$, thus 
the volume does not enter the sum \Ref{3dW1}; absorbing it in the
normalisation, together with the phase $e^{ - i\f{\pi}{4} - 4i \vartheta (j_0+\f12) }$,
we have
\equ\label{3dW2}
\ell_a^2 \, \ell_b^2 \, W_{1122}(T)
\simeq \f{1}{\cal N}\;  4 \, j_0^2 \, \lp^4 \sum_{j_1, j_2} \; \prod_e (2j_e+1) \;
\d j_1 \, \d j_2 \;e^{-\f{\alpha}{2}\sum_{i} \d j_i^2 
- \f{i}{2} \sum_{i,k} G_{ik} \d j_i \d j_{k}}.
\nequ
The sum can be easily performed by approximating it with a  Gaussian integral. 
In fact, in the limit $j\mapsto \infty$, $\lp\mapsto 0$, we can write
$$\sum_{j_1, j_2} \prod_e (2j_e+1)  \sim
(2j_0+1)^4 \int_0^\infty (2j_1+1) d j_1 \int_0^\infty (2j_2+1) d j_2
\simeq (2j_0)^6 \int dj_1 dj_2,$$ 
and also the factor $(2j_0)^6$ can be clearly absorbed in the normalisation.
Changing variables from $j_i$ to $z_i:=\d j_i$, we have $dj_i = dz_i$ and
\equ
\ell_a^2 \, \ell_b^2 \, W_{1122}(T)
\simeq \f{1}{\cal N}\; 4 \, j_0^2 \, \lp^4
\int dz_1 dz_2  \; z_1 z_2 \; e^{-\f12 z_i A_{ij} z_j},
\nequ
where $A_{ij} = \alpha \d_{ij} -i G_{ij}$.
At this point, we can fix the value of $\alpha$ to tune the correct vacuum
boundary state.
If we choose 
$
\alpha=\f4{3j_0},
$
which is consistent with \Ref{limit}, we have
\equ
A_{ij} = \alpha \left( \begin{array}{cc} 1+i \cot\theta & -\f i{\sin\theta} \\ \\
-\f i{\sin\theta} & 1+i \cot\theta \end{array}
\right).
\nequ
This can be easily recognised to be the kernel of an harmonic oscillator
(see the appendix), with frequency
\equ
\omega = \f\theta T \simeq \f83 \f1{j_0\lp} \equiv \f{2\alpha}{\lp}.
\nequ
Therefore, the vacuum boundary state together with the PR amplitude
reproduce in the semi--classical limit the kernel of an harmonic oscillator;
using the boundary geometry to evaluate $\ell_a^2 \, \ell_b^2 = (j_0\lp)^4$,
\equ\label{W3}
W_{1122}(T) \simeq  \f4{j_0^2} \, A^{-1}_{12} =
\f4{j_0^2} \, \f{e^{i\theta}}{2\alpha}=
\f8{j_0^2 \lp} \, \f{e^{i\om T}}{2\om}.
\nequ
The expression \Ref{W3} picks contributions from a single frequency. However,
suppose now that our tetrahedron is part of a discretisation of spacetime.
Then we expect to be able to sum over all frequencies admitted by the lattice, and
\Ref{W3} becomes
\equ
W_{1122}(T) = \f8{j_0^2 \lp} \,\sum_p  \f{e^{i\om_p T}}{2\om_p},
\nequ
which coincides with \Ref{W1122}, once we set $f_\lambda=8$, and we make
the usual lattice approximation $\int d^2p \sim \f1{(j_0\lp)^2}\sum_p $
(where $j_0\lp$ provides the characteristic length of the boundary geometry).

\section{Dependence on the distance}
To study the dependence on the spacetime distance $\ell$ of the 2-point function, it is convenient
to consider its absolute value. From \Ref{W3}, we read that at large scales 
\equ\label{large}
W(\ell):=|W_{1122}(\lp j_0)| = \f3{2\ell},
\nequ
thus reproducing the expected inverse power law of the 3d free graviton.

\begin{figure}[ht]
\begin{center}\includegraphics[width=8cm,angle=-90]{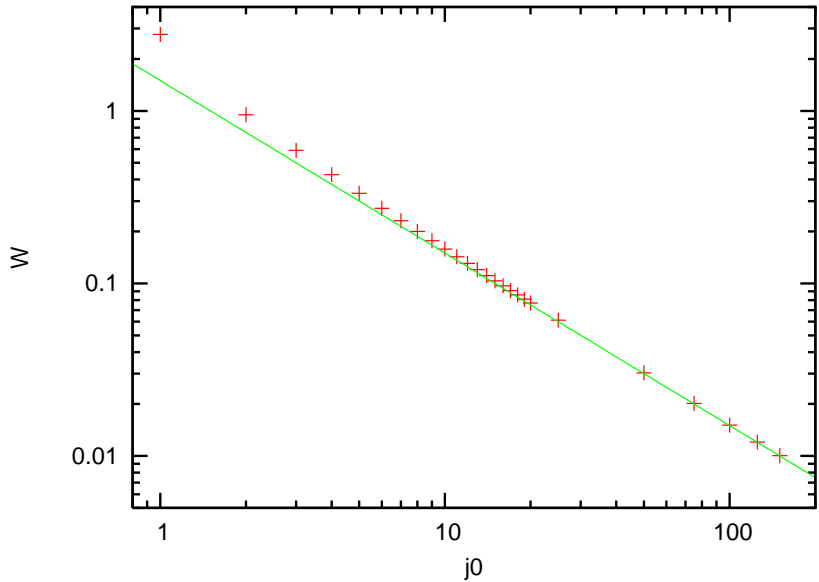}
\end{center}\caption{\small{The crosses are the numerical evaluations of \Ref{Wexact}, plotted
against the asymptotic behaviour $\f3{2j_0}$ on a bi--logarithmic scale. The agreement is very good for $j_0\geq 50$,
while deviations appear for smaller values.}}\label{fit}
\end{figure}

However, we can now use the exact expression \Ref{Wexact} to study the 
dependence on the distance at short scales. Notice that 
corrections to the inverse power law
are expected to arise at short scales; in the context of conventional QFT, these are
understood as graviton self--energies (see for instance  \cite{Donoghue}). In our approach,
they are indeed present, and come from higher order terms in the expansion \Ref{ReggeE}
as well as from departures from the asymptotic \Ref{asymp}. Here, we do not
attempt an analytical evaluation of these corrections, but simply briefly report a numerical study:
in Fig.\ref{fit} we plot some numerical values of \Ref{Wexact} when varying $j_0$. We see that
these are in very good agreement with the asymptotic behaviour \Ref{large} for $j_0\geq 50$.
For smaller values of $j_0$, we begin to see deviations from the inverse power law.
These deviations can be fitted by a series of inverse powers, which are the small scale
quantum corrections. For instance, allowing powers down to $-4$, we obtain the following fitting
function,
\eqa
W(\ell) &\simeq& \f3{2 \ell}[1 +\lp\f{0.67}{\ell} - \lp^2\f{0.74}{\ell^2} + \lp^3\f{0.92}{\ell^3}].  
\neqa

It should be added that these corrections are gauge--dependent quantities, just as the 
free propagator. In the conventional perturbative expansion of 3d GR, the self--energies vanish in the gauge
in which the free propagator vanishes, so that the full propagator is a pure gauge.
At the present stage of investigation, 
there is no reason to think that the spinfoam formalism should modify this picture,
thus we expect these corrections to be again pure gauge quantities; namely, we expect them to vanish in the same gauge
in which the free term vanishes.

\section{Conclusions}
We considered the recently appeared proposal for the extraction of the 2-point function
of linearised quantum gravity from the spinfoam formalism. To clarify some of the 
geometrical issues, we focused on a 3d toy model where spacetime is discretised by a single 
tetrahedron. We showed that, upon properly fixing the width of the Gaussian used as the
vacuum boundary state, the kernel for the space perturbations around flat space
behaves as the kernel of an harmonic oscillator, with frequency given by
$\f8{3 \ell}$, where $\ell=j_0\lp$ is the characteristic length of the boundary geometry.
Therefore, the propagator can be reconstructed as a collection of harmonic oscillators,
yielding the expected inverse power law in the large scale limit. 
Furthermore, we numerically studied the short scale behaviour, hinting that the $1/\ell$ 
behaviour is corrected by a series of higher inverse powers.
The proposal not only reproduces the expected large scale behaviour,
but it is also useful to study the short scale quantum corrections.

The main open issue is the extension of these results to many tetrahedra. On the one hand, this
is an important check of the stability of the result; on the other hand,
it would allow the study of the other components of the propagator, and
thus recover its full tensorial structure. 
In particular, notice that the full tensorial structure is needed to prove
in a more consistent way that the 2-point function computed here is just a pure gauge.

We hope that the clarifications here presented will help the calculations in the 4d case. 

\bigskip

\begin{center}{\bf Acknowledgments}\end{center}

I would like to thank Carlo Rovelli for encouragement and suggestions, John Baez for considerations
on the boundary Gaussian, Etera Livine for discussions, and Daniel Terno for help in using
Mathematica$^{\bigcirc {\hspace{-0.23cm}\scr R}}$, with which the numerical calculations have been done.

\appendix
\section{2-point function from the kernel for the harmonic oscillator}
Consider an harmonic oscillator, with mass $m=1$, in units $\hbar=1$,
and let us compute the 2-point function
$G(t_1, t_2) = \bra{0} x(t_1) x(t_2) \ket{0}$. 
This can be done in  the canonical formalism, giving
\equ\label{can}
\bra{0} x(t_1) x(t_2) \ket{0} = \bra{0} x e^{i H (t_2-t_1)} x \ket{0}
= \f1{2\om}e^{i\f32 \om T}, 
\nequ with $t_2-t_1:=T.$
However, it can also be computed starting from the propagation kernel
$W[x_1, x_2, T]$, as in \Ref{WGR}. We have
\equ
G(t_1, t_2) = \f1{\cal N} \int dx_1 \, dx_2 \,
W[x_1, x_2, T] \, \Psi_0[x_1]\, x_1 \, \Psi_0[x_2] \, x_2, 
\nequ
where the normalisation is  ${\cal N} = \int dx_1 \, dx_2 \,
W[x_1, x_2, T] \, \Psi_0[x_1] \, \Psi_0[x_2]$, and $\Psi_0[x]$
is the vacuum state.
Using the well known expressions (see for instance \cite{carlo})
\equ
\Psi_0[x] = \sqrt[4]{\f{\om}{\pi}}e^{-\f12\f{\om}{\hbar} x^2},
\qquad
W[x_1, x_2, T] = \sqrt{\f{\om}{2\pi i \sin\om T}} e^{-i\f{\om}{2}
\f{(x_1^2 + x_2^2)\cos\om T- 2 x_1 x_2}{\sin \om T}}, 
\nequ
we obtain
\equ\label{g}
G(t_1, t_2)  = \f1{\cal N} 
\int dx_1 \, dx_2 \,x_1\, x_2\, e^{-\f12 x_i A_{ij} x_j},
\nequ
with
\equ
A_{ij} = {\om} \left( \begin{array}{cc} 1+i \cot\om T & -\f i{\sin\om T} \\ \\
-\f i{\sin\om T} & 1+i \cot\om T \end{array} \right).
\nequ
The expression \Ref{g} is a Gaussian integral that can be readily evaluated, giving
\equ
G(t_1, t_2) = A_{12}^{-1} = \f1{2\om}e^{i \om T}.
\nequ
Up to the vacuum energy contribution $e^{i\f12\om T}$,
 the result coincides with the canonical evaluation \Ref{can}.

\end{document}